\documentclass{sf2a-conf2010}
\usepackage{graphicx}
\usepackage{hyperref}
\usepackage[]{natbib}

\begin{document}
\TitreGlobal{SF2A 2010}
%
\title{The catalog of radial velocity standard stars for the Gaia RVS: status and progress of the observations}
\author{L. Chemin}\address{Laboratoire d'Astrophysique de Bordeaux, UMR 5804 (CNRS, Universit\'e Bordeaux 1), 33271 Floirac Cedex, France}
\author{C. Soubiran$^1$}
\author{F. Crifo}\address{Observatoire de Paris, GEPI, UMR 8111 (CNRS , Universit\'e Denis Diderot Paris 7), 92195 Meudon, France}
\author{G. Jasniewicz}\address{GRAAL, UMR5024, (CNRS, Universit\'e Montpellier 2), 34095 Montpellier Cedex 05, France}
\author{L. Veltz}\address{Astrophysikalisches Institut Potsdam, Potsdam, An der Sternwarte 16, D-14482 Potsdam, Germany}
\author{D. Hestroffer}\address{Observatoire de Paris, IMCCE, UMR8028 (CNRS, Universit\'e Pierre \& Marie Curie Paris 6),  75014, Paris, France}
\author{S. Udry}\address{Observatoire de Gen\`eve, 51 Ch. des Maillettes, 1290 Sauverny, Switzerland}
\author{J. Berthier$^5$}
\author{A. Vigneron$^5$}
\author{D. Katz$^2$}
\author{A. Siebert}\address{Observatoire Astronomique de Strasbourg, UMR 7550 (CNRS, Universit\'e de Strasbourg),  67000, Strasbourg, France}
\runningtitle{The catalog of radial velocity standard stars for the Gaia RVS}
%
\setcounter{page}{237}
\index{L. Chemin}
\index{C. Soubiran}
\index{F. Crifo}
\index{G. Jasniewicz}
\index{L. Veltz}
\index{D. Hestroffer}
\index{S. Udry} 
\index{J. Berthier} 
\index{A. Vigneron} 
\index{D. Katz} 
\index{A. Siebert} 

\maketitle
\begin{abstract}
A new full-sky catalog of Radial Velocity standard stars is being built for the determination of 
the Radial Velocity Zero Point of the RVS on board of Gaia. After a careful selection of 1420 
candidates matching well defined criteria, we are now observing all of them to verify that 
they are stable enough over several years to be qualified as reference stars. We present the 
status of this long-term observing programme on three spectrographs : SOPHIE, NARVAL and CORALIE, 
complemented by the ELODIE and HARPS archives. Because each instrument has its own zero-point, we 
observe intensively IAU RV standards and asteroids to homogenize the radial velocity measurements. 
We can already estimate that ~8\% of the candidates have to be rejected because of 
variations larger than the requested level of 300 m s$^{-1}$.
\end{abstract}
\section{Introduction}
The purpose of this new spectroscopic catalog of standard stars is to calibrate the future radial velocities measured 
by the Radial Velocity Spectrometer (RVS) on board of the Gaia satellite \citep[see e.g. ][and references therein]{jas10}. 
We refer to \cite{cri09,cri10}  for a complete description of the selection criteria and of the 
ground observations of the 1420 candidates as reference stars.  

\section{Status of the observations}
\begin{figure}[ht!]
 \includegraphics[width=0.63\textwidth,viewport=1 1 700 419,clip]{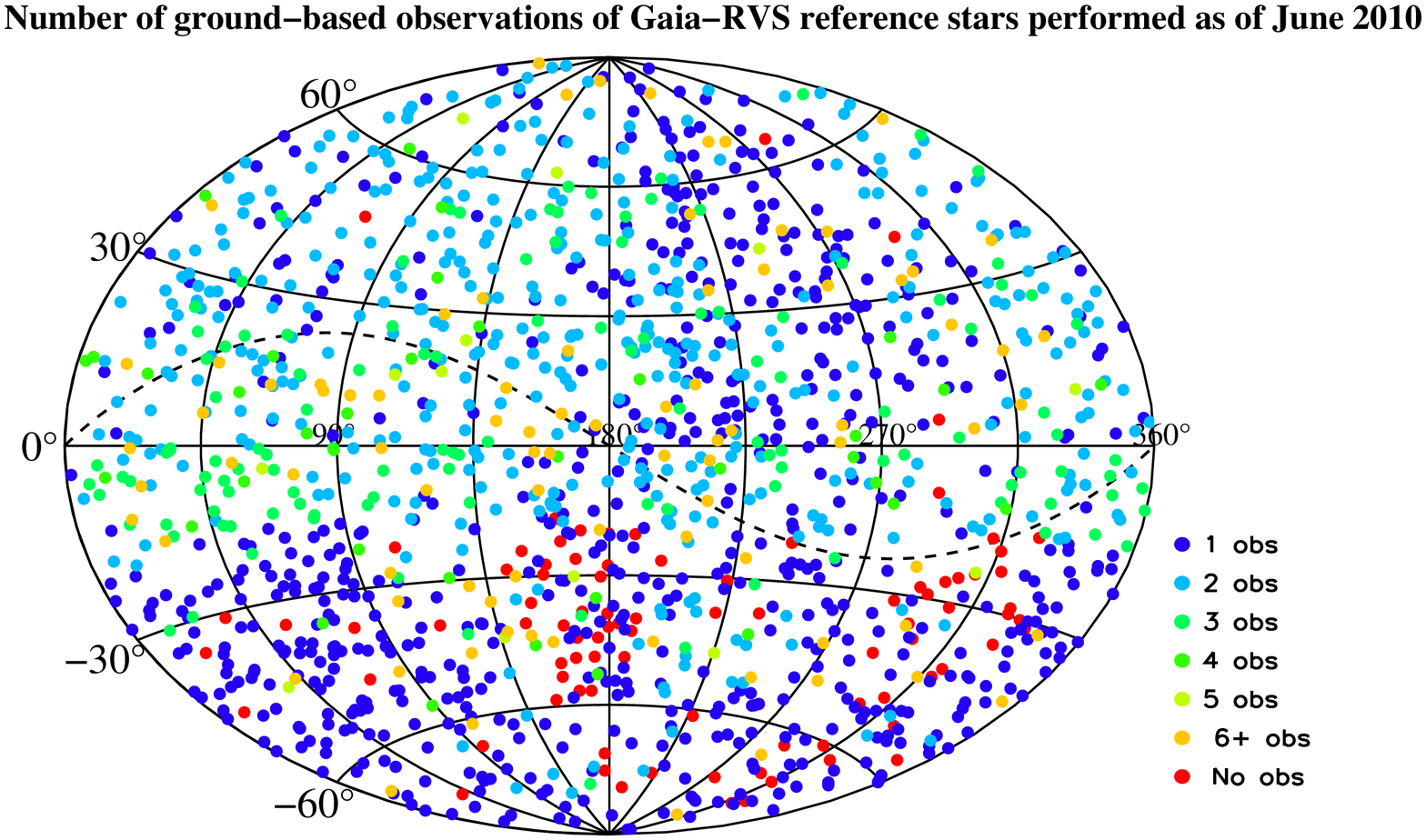} \includegraphics[width=0.37\textwidth]{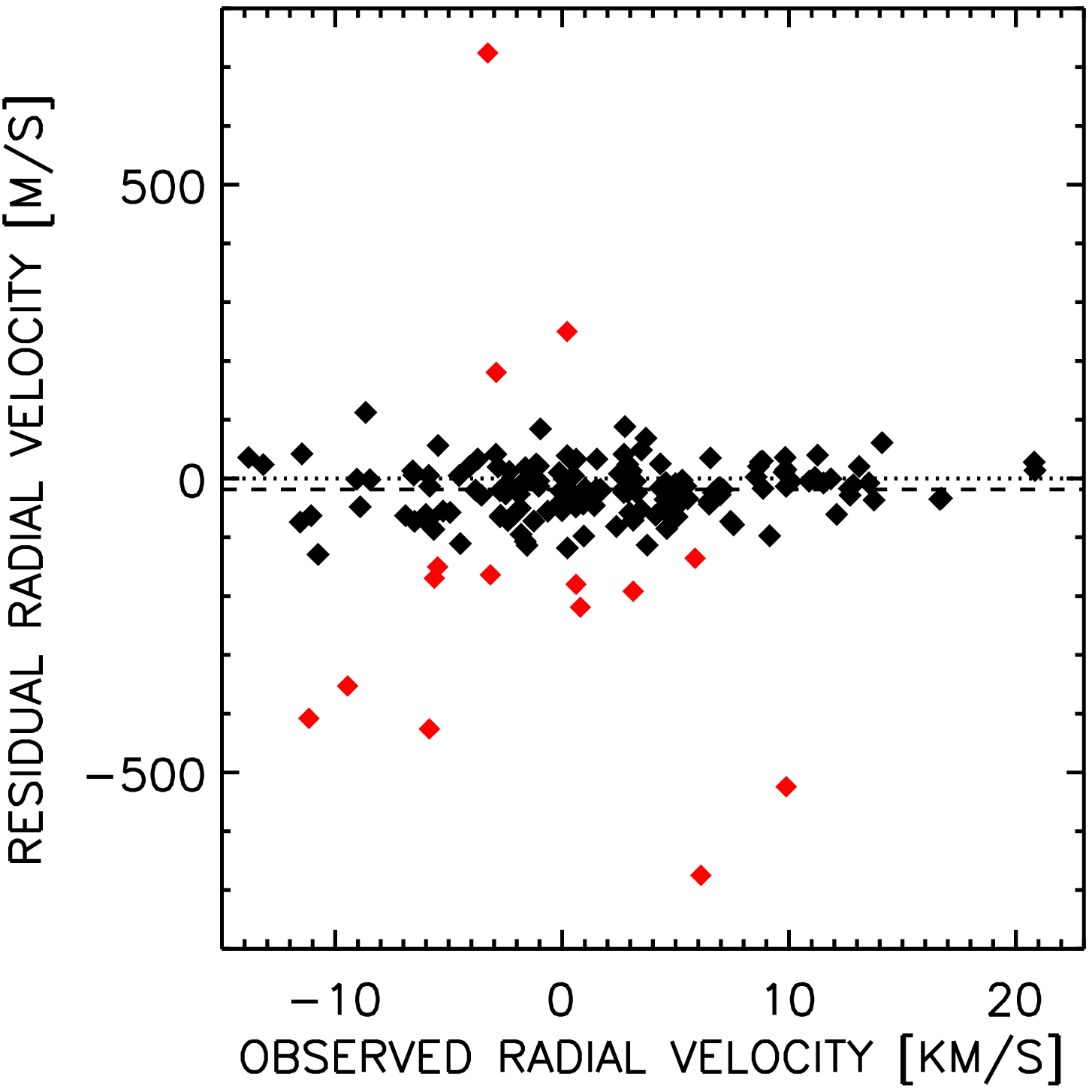}      
  \caption{\textbf{Left panel:} Number of ground-based observations of Gaia-RVS reference stars performed as of June 2010. The ecliptic is shown as a dashed line. 
  \textbf{Right panel:} Residual velocities (observed - theoretical) for asteroids as a function of their observed  velocities. The red dots deviate by more than $3\sigma$.}
  \label{chemin:fig1}
\end{figure}

A total of 4035 measurements   is currently available for 1330 stars. It consists in new  and archived observations performed  
with the NARVAL (98 measurements), CORALIE (688),  SOPHIE (902), ELODIE (1057) and HARPS (1290) high-resolution spectrographs. 
Figure~\ref{chemin:fig1} (left panel) represents the spatial distribution in the equatorial frame of the number of measurements already obtained for the 1420 candidates.
The majority of stars still lacking observations is located in the Southern part of the sky, because the Southern programme on CORALIE started later. 
 In the North ($\delta > -15^\circ$), $\sim$200 stars still lack a second measurement. The Northern programme should be completed in 2011. 
   
\section{Preliminary results}
\begin{itemize}
\item \textit{\textbf{Radial velocities of stars:}}
We have first compared the radial velocities of 320 stars we have in common with \cite{nid02}.  A mean difference of $-40$ m s$^{-1}$ exists between both studies. 
This illustrates the zero-point issue that has to be solved for the calibration of the RVS \citep{jas10}.  
Among those 320 stars 27 objects deviate by more than $\sim$300 m s$^{-1}$.  
Such  a discrepancy between both studies may be due to variable stars that should not be considered as standard objects in a future analysis.

We have also done a preliminary statistical analysis our catalog. When selecting a sub-sample of 673 candidates for which at least two   measurements have been performed 
it is seen that the velocity for $\sim$72\% of them does not vary by more than 100 m s$^{-1}$ during a time baseline of 0.5-2 years. 
Such a stability of radial velocities is very important to get the most accurate calibration of the RVS. 
The time variability of the catalog will be studied during the whole lifetime of the Gaia mission.
Notice that $\sim$8\% of the 673 stars exhibit a velocity variation of more than 300 m s$^{-1}$.  Those objects likely correspond to variable stars.  

\item \textit{\textbf{Radial velocities of asteroids:}}
Observations of asteroids are very important for this project because they will allow the derivation of the zero point of the radial velocities for all reference sources. 
171 measurements of  70 asteroids have been performed until now with SOPHIE. Their velocities have been compared with the theoretical values (Fig.~\ref{chemin:fig1}, right panel)
 which have been  derived using the  MIRIADE webservice of the virtual observatory at IMCCE. 
The scatter of the residual (observed minus calculated) velocity is $\sigma \sim 45$ m s$^{-1}$. 
Points that are more deviant than $3\sigma$ (red symbols) correspond all to low signal-to-noise observations due to bad transparency conditions, or to badly derived velocity centroids due to e.g. 
double peaks in the cross-correlation function (observing conditions, contamination by the moon, ...). 
We are currently investigating the correlations of the observed and computed velocities with the physical properties of the asteroids (diameter, shape, rotation, phase, etc).
\end{itemize}

\begin{acknowledgements}
We are very grateful to the AS-Gaia, the PNPS and PNCG  for the financial support and help in this project.
\end{acknowledgements}

\end{document}